\documentclass{nature}

\usepackage{amssymb} %% two column, final layout
\usepackage[implicit=false]{hyperref}
\usepackage{bbm}
\usepackage{mathrsfs}
\usepackage{amsfonts}
\usepackage{graphicx}
\usepackage{cite}
\usepackage[linesnumbered,ruled,lined]{algorithm2e}
\usepackage{array}
\usepackage{subfigure}
\usepackage{multirow}
\usepackage{color}
\usepackage{amsmath}
\usepackage{bm}
\usepackage{upgreek}
\usepackage{float}

\usepackage{authblk}
\usepackage{booktabs}  
\usepackage{threeparttable} 

\usepackage{amssymb}
\usepackage{pifont}

\usepackage{url}

\title{Large-scale phase retrieval}

\author[]{Xuyang Chang,$^{1,2}$ Liheng Bian,$^{1,2,*}$, and Jun Zhang$^{1,2}$
}

\begin{document}

\maketitle

\begin{affiliations}
 \item School of Information and Electronics, Beijing Institute of Technology, Beijing 100081, China
 \item Advanced Research Institute of Multidisciplinary Science, Beijing Institute of Technology, Beijing 100081, China\\
 $^*$bian@bit.edu.cn
\end{affiliations}

\begin{abstract}
High-throughput computational imaging requires efficient processing algorithms to retrieve multi-dimensional and multi-scale information. In computational phase imaging, phase retrieval (PR) is required to reconstruct both amplitude and phase in complex space from intensity-only measurements. The existing PR algorithms suffer from the tradeoff among low computational complexity, robustness to measurement noise and strong generalization on different modalities. In this work, we report an efficient large-scale phase retrieval technique termed as \emph{LPR}. It extends the plug-and-play generalized-alternating-projection framework from real space to nonlinear complex space. The alternating projection solver and enhancing neural network are respectively derived to tackle the measurement formation and statistical prior regularization. This framework compensates the shortcomings of each operator, so as to realize high-fidelity phase retrieval with low computational complexity and strong generalization. We applied the technique for a series of computational phase imaging modalities including coherent diffraction imaging, coded diffraction pattern imaging, and Fourier ptychographic microscopy. Extensive simulations and experiments validate that the technique outperforms the existing PR algorithms with as much as 17dB enhancement on signal-to-noise ratio, and more than one order-of-magnitude increased running efficiency. Besides, we for the first time demonstrate ultra-large-scale phase retrieval at the 8K level (7680$\times$4320 pixels) in minute-level time.
\end{abstract}

\newpage
Wide field of view and high resolution are both desirable for various imaging applications, such as medical imaging\cite{pahlevaninezhad2018nano,lombardini2018high,zheng2013wide,fan2019video} and  remote sensing\cite{wang2011space}, providing multi-dimensional and multi-scale target information. As the recent development of computational imaging, large-scale detection has been widely employed in a variety of computational imaging modalities \cite{brady2012multiscale, zheng2013wide, wang2016computational, fan2019video}. These computational imaging techniques largely extend the spatial-bandwidth product (SBP) of optical systems from million scale to billion scale. As an example, the SBP of the RUSH microscope platform \cite{fan2019video} has reached as high as $1.7 \times 10^{8}$. Such large amount of data poses great challenge for post software processing. Therefore, large-scale processing algorithms with low computational complexity and high fidelity are of great significance for those imaging and perception applications in various dimensions \cite{yuan2020plug}.

In computational phase imaging, phase retrieval (PR) is required to reconstruct both amplitude and phase in complex space from intensity-only measurements. This problem originates from the limitation of low response speed of photodetectors that impedes direct acquisition of light wavefront. Mathematically, the underlying goal of PR is to estimate an unknown complex-field signal from the intensity-only measurements of its complex-valued transformation, which is described as
\begin{equation}
\label{eq:PR}
{I}=|\boldsymbol{A} u|^2+\omega,
\end{equation} 
where $u$ is the underlying signal to be recovered $\left(u \in \mathbb{C}^{n \times 1}\right)$, $I$ contains the intensity-only measurements $\left(I \in \mathbb{R}^{m \times 1}\right)$, $ \boldsymbol{A}$ represents measurement matrix $\left(\boldsymbol{A} \in \mathbb{R}^{n \times n} \text { or } \mathbb{C}^{n \times n}\right)$, and $\omega$ stands for measurement noise. Phase retrieval has been widely applied in plenty fields such as astronomy, crystallography, electron microscopy and optics \cite{shechtman2015phase}. It solves various nonlinear inverse problems in optical imaging, such as coherent diffraction imaging \cite{miao1999extending}(CDI), coded diffraction pattern imaging \cite{candes2015phase}(CDP), Fourier ptychographic microscopy \cite{zheng2013wide} (FPM) and imaging through scattering medium\cite{katz2014non}.

In the past few decades, different phase retrieval algorithms have been developed. Gerchberg and Saxton pioneered the earliest alternating projection (AP) algorithm in the 1970s \cite{gerchberg1972practical}, which was then extended by Fienup et al. with several variants \cite{fienup1982phase}. Due to its strong generalization ability, AP has been widely employed in multiple phase imaging models. Nevertheless, it is sensitive to measurement noise, suffering from poor noise robustness. Afterwards, researchers introduced optimization into PR, deriving a series of semi-definite programming (SDP) based algorithms \cite{candes2013phaselift, vandenberghe1996semidefinite} and Wirtinger flow (WF) based algorithms \cite{candes2015phasewf, chen2015solving, zeng2020coordinate}. These techniques enhances robustness to measurement noise, but they require high computational complexity and high sampling rate, making them inapplicable for large-scale phase retrieval. Although the sparsity prior of natural images in transformed domains can be incorporated as an additional constraint to lower sampling rate \cite{katkovnik2017phase, metzler2016bm3d, metzler2018prdeep}, it further increases computational complexity.

Recently, the booming deep learning (DL) technique has also been introduced for phase retrieval\cite{rivenson2018phase}. Following the large-scale training framework, the DL strategy outperforms the above traditional PR techniques with higher fidelity. However, it provides poor generalization that each suits only for specific models, such as holography \cite{rivenson2018phase} and FPM \cite{kappeler2017ptychnet}. For different models and even different system parameters, the deep neural network requires to be retrained with new large-scale data sets. To sum, despite of different workflows, the above existing PR algorithms suffer from the tradeoff among low computational complexity, robustness to measurement noise and strong generalization, making them inapplicable for large-scale phase retrieval.

\begin{figure}[h]
  \centering
  \includegraphics[width=0.83\linewidth]{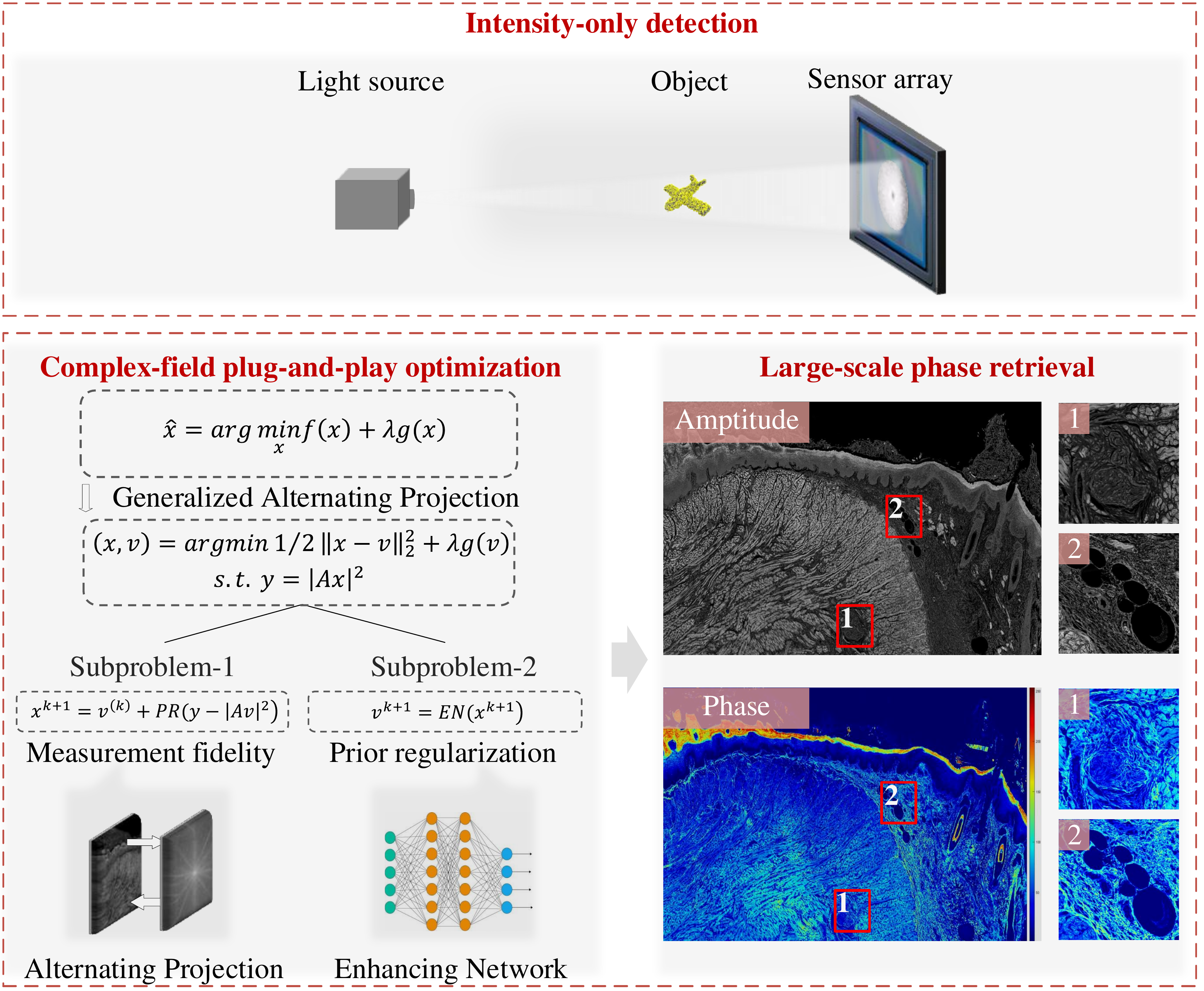}
  \caption{{{\bf The schematic of the reported \emph{LPR} technique for large-sacle phase retrieval.} \emph{LPR} decomposes the large-scale phase retrieval problem into two subproblems under the PNP-GAP framework, and introduces the efficient alternating projection (AP) and enhancing network solvers for alternating optimization. The workflow realizes robust phase retrieval with low computational complexity and strong generalization on different imaging modalities.}}
  \label{fig:schematic}
\end{figure}

In this work, we report an efficient large-scale phase retrieval technique termed as \emph{LPR}, as sketched in Fig. \ref{fig:schematic}. It builds on the plug-and-play (PNP) \cite{venkatakrishnan2013plug} optimization framework, and extends the efficient generalized-alternating-projection (GAP) \cite{yuan2020plug, liao2014generalized, yuan2016generalized} strategy from real space to nonlinear complex space. The complex-field PNP-GAP scheme ensures strong generalization of \emph{LPR} on various imaging modalities, and outperforms the conventional first-order PNP techniques (such as ISTA \cite{bioucas2007new} and ADMM \cite{venkatakrishnan2013plug}) with fewer auxiliary variables, lower computational complexity and faster convergence. As PNP-GAP decomposes reconstruction into separate sub-problems including measurement formation and statistical prior regularization \cite{liu2018rank, yuan2020plug}, we further introduce an alternating projection solver and an enhancing neural network respectively to solve the two sub-problems. These two solvers compensate the shortcomings of each other, allowing the optimization to bypass the poor generalization of deep learning and poor noise robustness of AP. As a result, \emph{LPR} enables generalized large-scale phase retrieval with high fidelity and low computational complexity, making it a state-of-the-art method for various computational phase imaging applications. 

We compared \emph{LPR} with the existing PR algorithms on extensive simulation and experiment data of different imaging modalities. The results validate that compared to the AP based PR algorithms, \emph{LPR} is robust to measurement noise with as much as 17dB enhancement on signle-to-noise ratio. Compared with the optimization based PR algorithms, the running time is significantly reduced by more than one order of magnitude. Finally, we for the first time demonstrated ultra-large-scale phase retrieval at the 8K level (7680$\times$4320 pixels) in minute-level time, where most of the other PR algorithms failed due to unacceptable high computational complexity.

\section*{Results}
We applied \emph{LPR} and the existing PR algorithms on both simulation and experiment data of three computational phase imaging modalities including CDI, CDP and FPM, to investigate respective pros and cons. The competing algorithms for comparison includes the alternating projection technique (AP)\cite{gerchberg1972practical,fienup1982phase}, the SDP based techniques (PhaseMax (PMAX)\cite{goldstein2018phasemax}, PhaseLift (PLIFT)\cite{candes2013phaselift}, PhaseLamp (PLAMP)\cite{dhifallah2017phase}), the Wirtinger flow based techniques (Wirtinger Flow (WF)\cite{candes2015phasewf}, Reweighted Wirtinger Flow (RWF)\cite{yuan2017phase}), the amplitude flow based techniques\cite{wang2017solving,wang2018phase} (AmpFlow (AF), Truncated AmpFlow (TAF), Reweighted AmpFlow (RAF)), Coordinate Descent (CD)\cite{zeng2017coordinate}, KACzmarz (KAC)\cite{wei2015solving} and the deep learning based prDeep technique\cite{metzler2018prdeep}. All the algorithm parameters were tuned based on the Phasepack \cite{chandra2019phasepack} for respective best performance. The convergence is determined when the intensity difference of reconstructed image between two successive iterations is smaller than a preset threshold. We employed the peak signal-to-noise ratio (PSNR) and structural similarity index (SSIM) \cite{wang2004image} to quantify reconstruction quality. All the calculation was tested on a desktop PC with an Intel i7-9700 CPU, 16G RAM and an Nvidia GTX 1660s GPU.

\paragraph{Coherent diffraction imaging.}
CDI is a representative non-interferometric phase imaging technique, and has been widely applied in physics, chemistry and biology due to its simple setup \cite{shechtman2015phase}. It illuminates a target using coherent plane wave, and records the intensity of far-field diffraction pattern. By oversampling the diffracted light field and applying phase retrieval, both the target's amplitude and phase information can be reconstructed. Mathematically, the measurement formation of CDI is
\begin{equation}
\label{eq:CDI}
I = |\mathcal{F}(u)|^{2},
\end{equation} 
where $u$ denotes the target information, and $\mathcal{F}$ represents the Fourier transformation that approximates the far-field diffraction.

Following the above formation model, we employed a high-resolution image (1356$\times$2040 pixels) from the DIV2K\cite{agustsson2017ntire} dataset as the latent signal to synthesize CDI measurements. Due to the uniqueness guarantee of solution, CDI requires at least 4 times oversampling in the Fourier domain\cite{miao2015beyond}. Correspondingly, we padded zeros around the image matrix to generate a 2712$\times$4080 image. %The location of orignial image regards as a prior information.
We implemented Fourier transform to the image and retained only its intensity as measurements. Additionally, to investigate the techniques' robustness to measurement noise, we further added different levels of white Gaussian noise (WGN) to the measurements.

Table \ref{tab:1} presents the quantitative reconstruction evaluation of different techniques. The results show that the CD and KAC methods failed with no convergence. This is because these techniques require higher sampling ratio. The PLIFT and PLAMP methods do not work as well, because they require matrix lifting and involve higher dimensional matrix that is out of memory in large-scale reconstruction. The other methods except for prDeep obtain little improvement compared to the AP algorithm. Specifically, the WF, AF and PMAX methods even degrade due to limited sampling ratio and noise corruption. The reconstruction of prDeep is better than the conventional algorithms, but with only 2dB enhancement on PSNR, and almost no SSIM improvement compared to AP. In contrast, \emph{LPR} produces significant enhancement on reconstruction quality, with as much as 6dB and 0.29 improvement on PSNR and SSIM, respectively. Due to limited space, the detailed visual comparison of different techniques is presented in Fig. S1 (supplementary information), which coincides with the above quantitative results. 

Table \ref{tab:1} also presents the running time of these techniques. Because all the other algorithms used the result of AP as initialization, we recorded the excess time as the running time of these algorithms. From the results, we can see that prDeep consumes the most running time. \emph{LPR} takes the same level of running time compared to the conventional algorithms, but with significantly improved reconstruction quality. 

\begin{figure*}[h]
  \centering
  \includegraphics[width=1.02\linewidth]{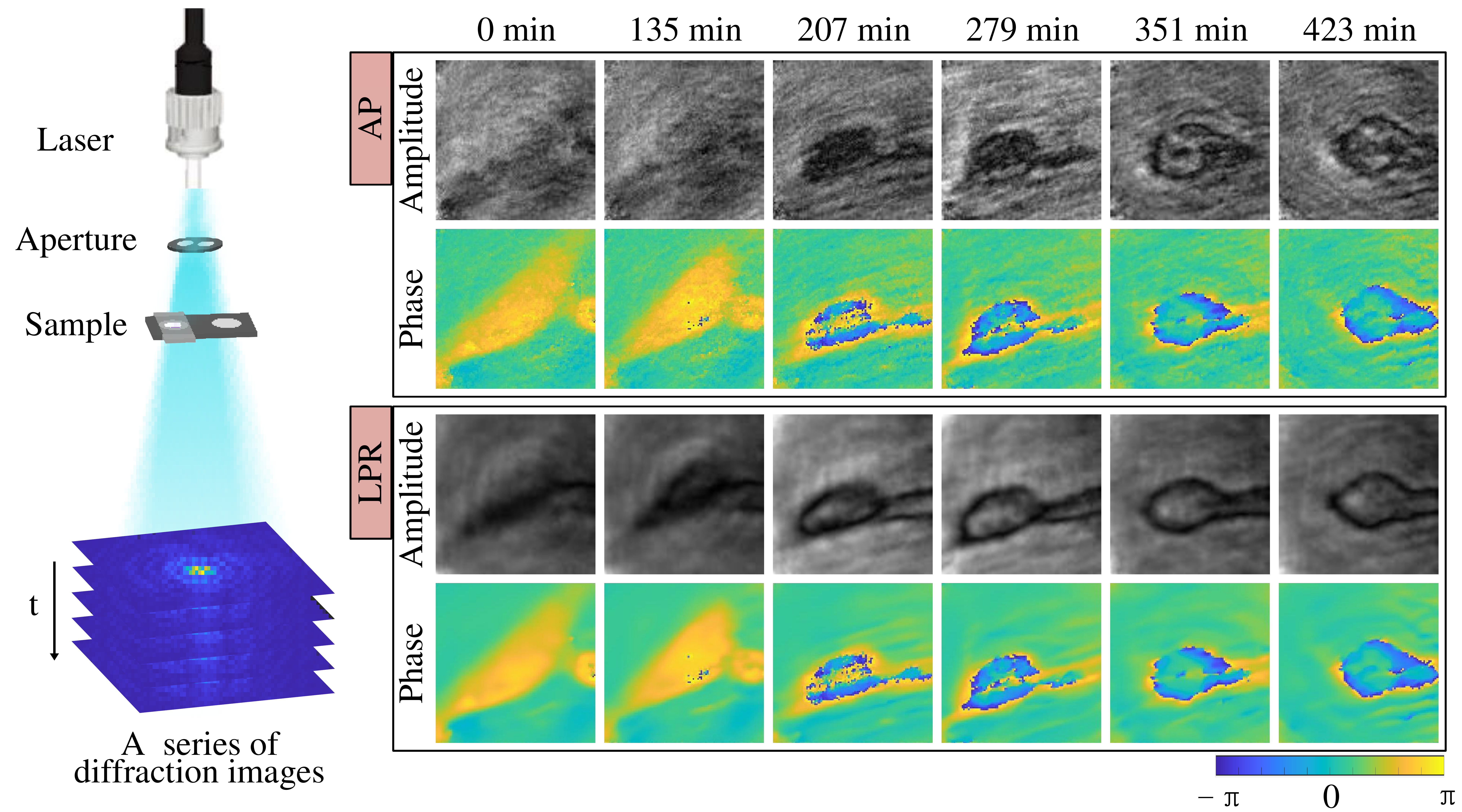}
  \caption{{{\bf Comparison of experiment results under the CDI modality\cite{lo2018situ}.} A dual-pinhole aperture is illuminated by a coherent light. A live glioblastoma cell sample is imaged in a time series of diffraction patterns. The reconstructed results describe the fusion process of two glioblastoma cells and form a high-density area. The AP technique is sensitive to measurement noise, and produces unsatisfying results. The reported \emph{LPR} technique enables to remove noise artifacts and preserve fine details with high fidelity.}}
  \label{fig:CDIreal}
\end{figure*}

We further compared these algorithms on experiment CDI data\cite{lo2018situ}, to validate their effectiveness in practical applications. The imaging sample is live glioblastoma cell line U-87 MG. The setup includes a HeNe laser (543nm 5mW), a dual pinhole aperture which consists of two 100 µm pinholes spaced 100 µm apart from edge to edge, a 35 mm objective lens and a CCD camera (1340$\times$1300, 16 bits). The sequential measurements contain far-field diffraction patterns of several moments in the cell fusion process. Because the conventional algorithms obtain little improvement compared to AP and prDeep is not applicable for complex-field sample \cite{metzler2018prdeep}, we only present the reconstruction results of AP and \emph{LPR} in Fig. \ref{fig:CDIreal}. The results show that there exists serious noise artifacts in AP reconstruction, especially in the amplitude images. The cells are almost submerged by background noise at 0 and 135 min, and the contours and edges of cells can not be clearly observed. In comparison, \emph{LPR} produces high-fidelity results that effectively preserve fine details while attenuating measurement noise. The complete results of all the 48 moments are shown in Fig. S2 - Fig. S5 (supplementary information).

\paragraph{Coded diffraction pattern imaging.}
CDP \cite{candes2015phase} is a coded version of CDI, which introduces wavefront modulation to increase observation diversity. The strategy of multiple modulations and acquisitions enables to effectively bypass the oversampling limitation of the conventional CDI. 
Generally, the target light field is additionally modulated by a spatial light modulator (SLM), and the measurements after far-field Fraunhofer diffraction can be modeled as
\begin{equation}
	\label{eq:CDP}
I=|\mathcal{F}(u\odot d)|^{2},
\end{equation}
where $d$ represents the modulation pattern, and $\odot$ denotes the Hadamard product.

We simulated CDP measurements with five and single phase modulations, respectively. The modulation patterns $d$ are subject to Gaussian distribution \cite{candes2015phase}. 
We employed the same image as CDI to be the ground-truth signal, and added various levels of WGN to the measurements. Table \ref{tab:2} presents the quantitative evaluation of different techniques under the CDP modality (5 modulations). The results show that the Wirtinger flow based techniques (WF and RWF) failed because of insufficient measurements\cite{candes2015phasewf}. The PLIFT and PLAMP methods are still out of memory. The other conventional methods produce either little improvement or even worse reconstruction compared to AP. Although prDeep outperforms AP, it consumes around triple running time with high computational complexity. In comparison, the reported \emph{LPR} obtains the best  reconstruction performance, with as much as 8.3dB on PSNR and 0.61 on SSIM. Besides, it also shares the same level of running time as AP, which maintains the highest efficiency among all the algorithms. The detailed visual comparison of different methods is presented in Fig. S6 (supplementary information). 

\begin{figure*}[h]
  \centering
  \includegraphics[width=\linewidth]{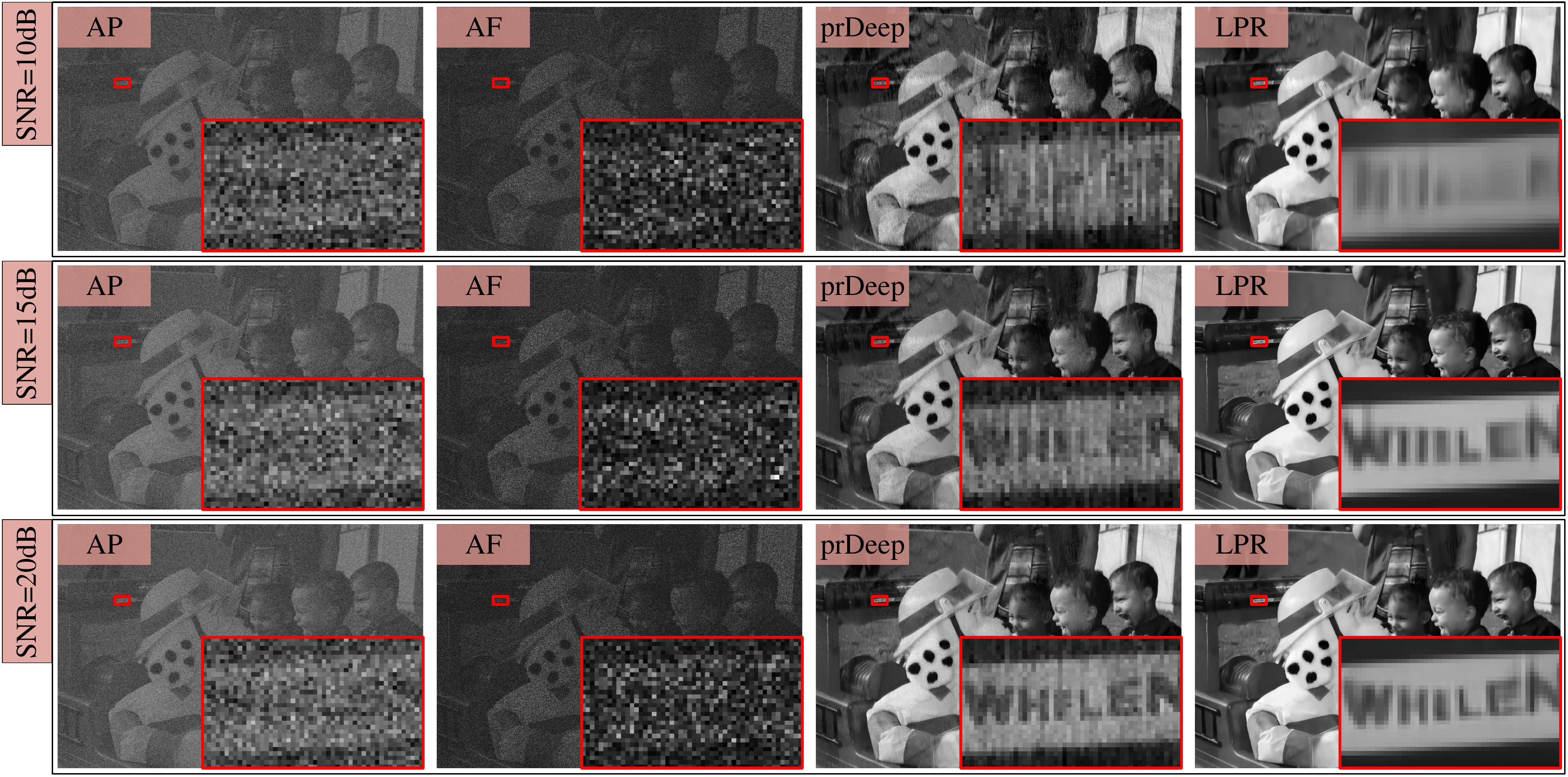}
  \caption{{{\bf Visual comparison under the CDP imaging modality (single modulation).} In such a low sampling ratio with measurement noise, all the  conventional algorithms produce low-contrast resolution. The prDeep technique also produces serious reconstruction artifacts. The reported  \emph{LPR} technique outperforms the other methods with much higher fidelity.}}
  \label{fig:CDP1}
\end{figure*}

To further demonstrate the strong reconstruction performance of \emph{LPR}, we also compared these algorithms in the case of limited sampling ratio with only single modulation, as shown in Tab. \ref{tab:3} and Fig. \ref{fig:CDP1}. Due to extremely insufficient measurements, most of the methods failed with either no convergence or poor reconstruction quality. Under heavy measurement noise, the target information is either buried or smoothed. In contrast, the reported \emph{LPR} thchnique enables as much as 17dB enhancement on PSNR and 0.8 improvement on SSIM. As validated by the colse-ups in Fig. \ref{fig:CDP1}, \emph{LPR} is able to retrieve fine details, even in the case of heavy measurement noise. Meantime, it is effective to attenuate noise and artifacts, producing smooth background.

\paragraph{Fourier ptychographic microscopy.}
FPM is a novel technique to increase optical system's bandwidth  for wide-field and high-resolution imaging. It illuminates the target with coherent light at different incident angles, and acquires corresponding images that contain information of different sub-regions of the target's spatial spectrum. Mathematically, the measurement formation model of FPM is
\begin{equation}
\label{eq:FPM}
I=\left|\mathcal{F}^{-1}[P \odot \mathcal{F}\{u \odot \mathcal{S}\}]\right|^{2},
\end{equation}
where $\mathcal{F}^{-1}$ is inverse Fourier transform, $P$ denotes system's pupil function, and $\mathcal{S}$ represents the wave function of incident light.

Following the formation model, we first implemented a simulation comparison with the following setup parameters: the wavelength is 625nm, the numerical aperture (NA) of objective lens is 0.08, the height from the light source to the target is 84.8mm, and the distance between adjacent light sources is 4mm. The pixel size of camera is 3.4$\mu$m. Two microscopy images of blood cells \cite{blood_cell} (2048$\times$2048 pixels) were employed as the latent high-resolution (HR) amplitude and phase, respectively. The size of captured low-resolution images (LR) was one fourth of the HR images.

\begin{figure*}[h]
  \centering
  \includegraphics[width=\linewidth]{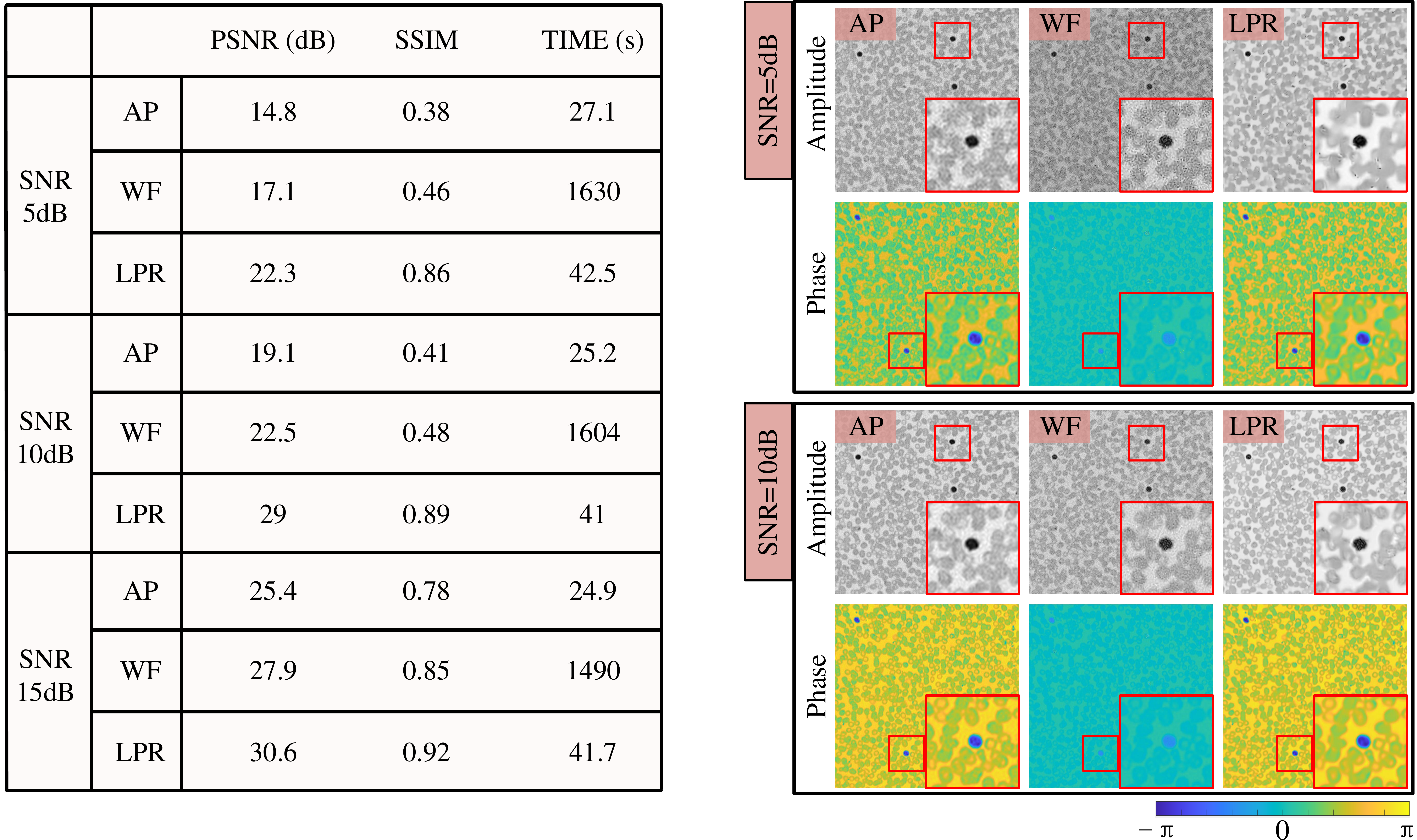}
  \caption{{{\bf Comparison of simulation results under the FPM modality.} The left table presents quantitative comparison, while the right images show visual comparison. AP suffers from poor noise robustness. WF requires high computational complexity with longer running time (more than one order of magnitude). In contrast, \emph{LPR} produces the highest reconstruction quality with as much as nearly 10dB enhancement on PSNR (SNR = 10), and consumes the same order of running time as AP.}}
  \label{fig:FPMsimulation}
\end{figure*}

Figure \ref{fig:FPMsimulation} presents the reconstruction results of AP \cite{zheng2013wide}, WF \cite{bian2015fourier} and \emph{LPR}. AP is sensitive to measurement noise. WF can better handle noise, but it requires high computational complexity and long running time (more than one order of magnitude). Compared with AP, \emph{LPR} obtains as much as nearly 10dB enhancement on PSNR (SNR  = 10). Besides, it consumes the same order of running time as AP. The visual comparison also validates that \emph{LPR} enables high-fidelity reconstruction of both amplitude and phase. Due to space limitation, we present another set of simulation results in Fig.  S7 (supplementary information).

\begin{figure*}[h]
  \centering
  \includegraphics[width=\linewidth]{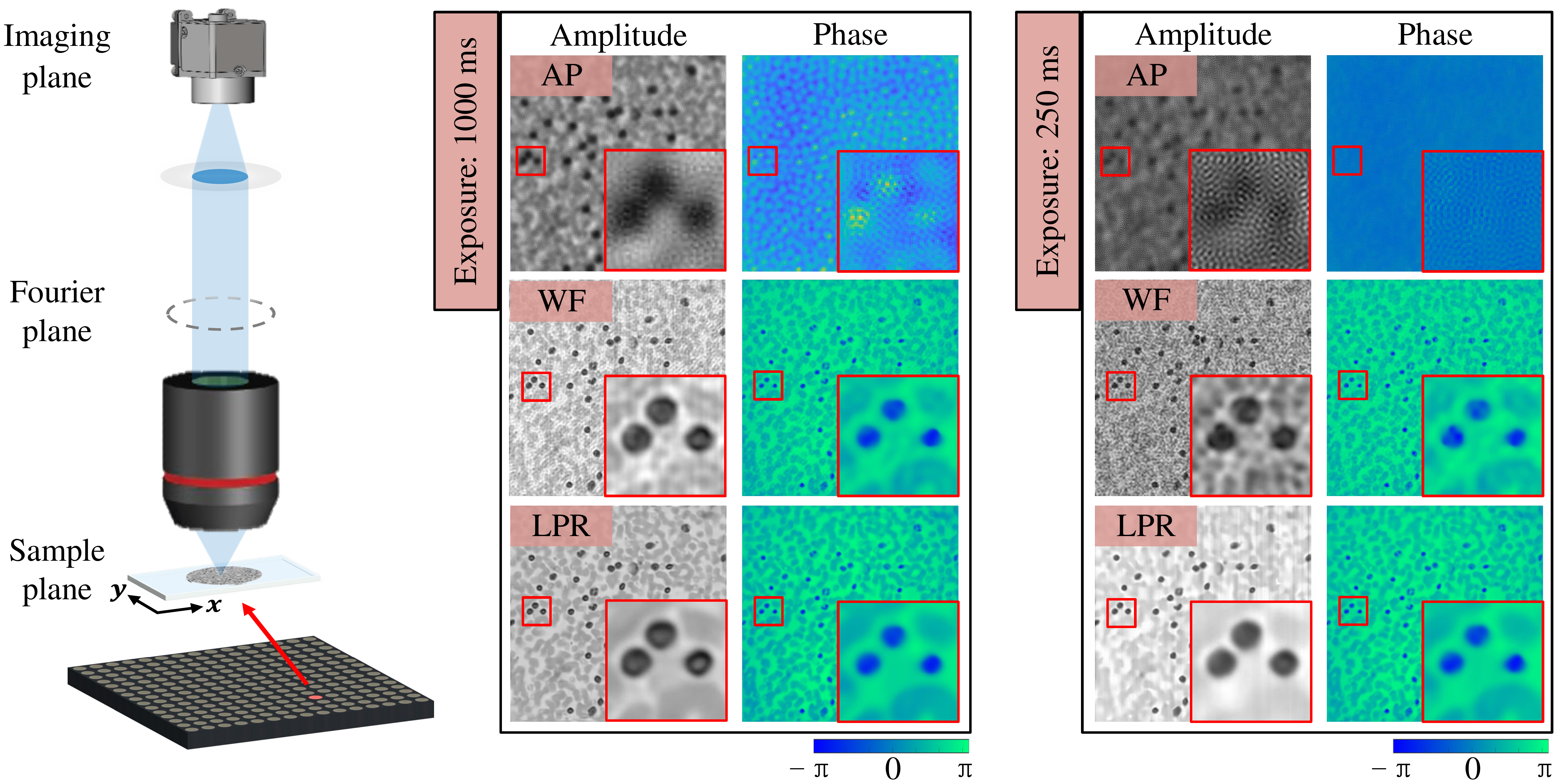}
  \caption{{{\bf 
Comparison of experiment results under the FPM modality.} The target is a red blood cell sample that is prepared on a microscope slide stained with Hema 3 stain set (Wright-Giemsa). The limited exposure results in serious measurement noise, which directly flows into the reconstruction results of AP. The WF technique outperforms AP, but it still degrades a lot under short exposure time (250ms). The reported \emph{LPR} technique maintains strong robustness to measurement noise, and enables to retrieve clear cell structure and morphology details.}}
  \label{fig:FPMreal}
\end{figure*}

We also implemented the algorithms on experiment FPM measurements. The imaging sample is blood smear stained by HEMA 3 Wright-Giemsa. The setup consists of a 15$\times$15 LED array, a 2$\times$ 0.1 NA objective lens (Olympus), and a camera with 1.85$\mu$m pixel size. The central wavelength of the LEDs is 632nm, and the lateral distance between adjacent LEDs is 4mm. The LED array is placed 80mm from the sample. We captured two sets of 225 LR images that  correspond to the 15$\times$15 LEDs, respectively under 1000ms and 250ms exposure time. The reconstructed results are presented in Fig. \ref{fig:FPMreal}, which show that AP is seriously degraded under limited exposure. Only the cell nucleus can be observed in amplitude, and other details are lost. \emph{LPR} produces state-of-the-art reconstruction performance. The measurement noise is effectively removed, and the cell structure and morphology details are clearly retrieved.

\begin{figure*}[h!]
  \centering
  \includegraphics[width=\linewidth]{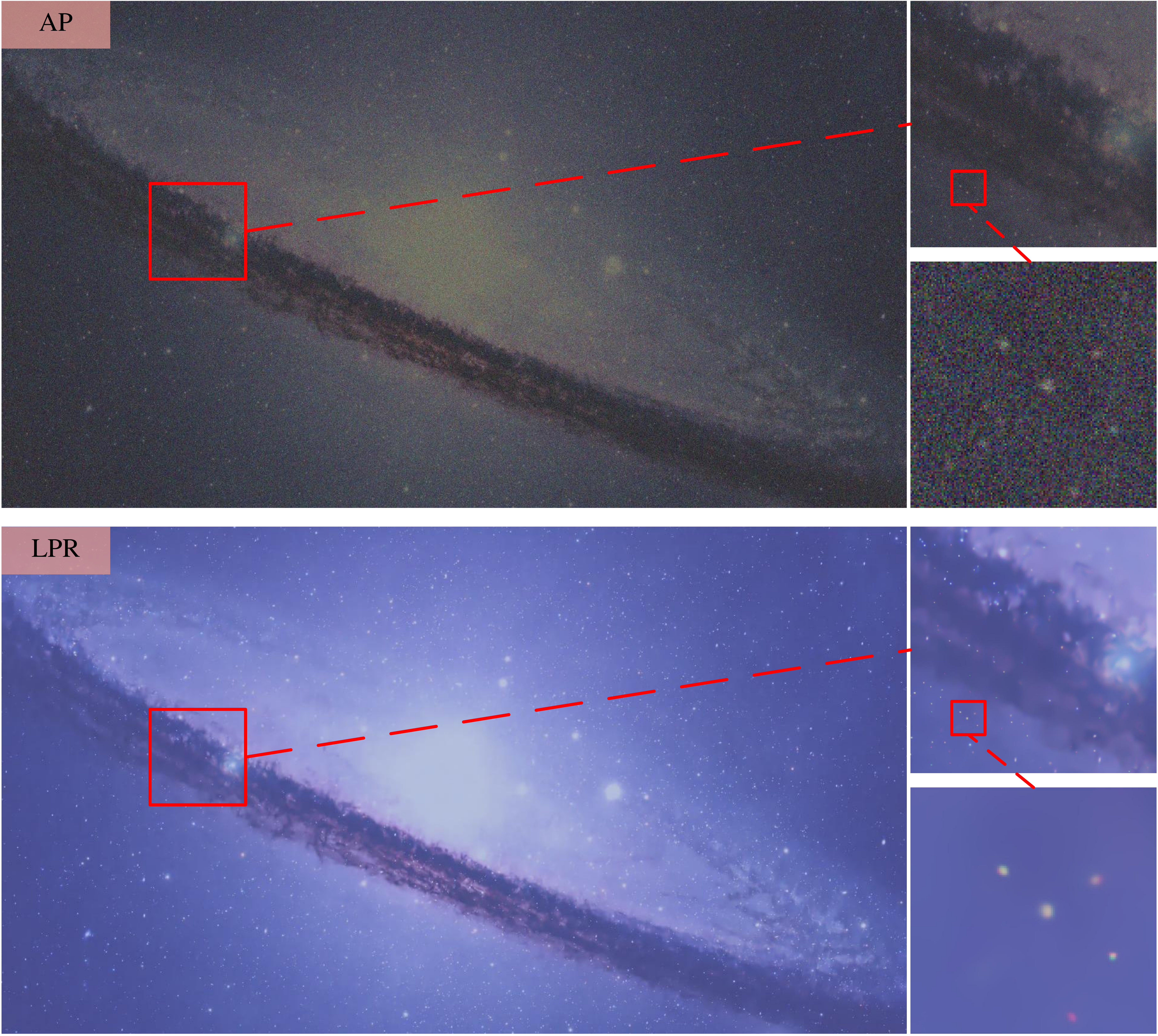}
  \caption{{{\bf 
The first demonstration of ultra-large-scale phase retrieval at the 8K level (7680$\times$4320$\times$3 pixels).} The imaging modality is CDP with 5 modulations. At such a large scale, only the AP and the reported \emph{LPR} techniques still work, while the other ones fail due to high compulational complexity. The results validate that \emph{LPR} significantly outperforms AP with effective noise removal and detail reservation.}}
  \label{fig:ultra}
\end{figure*}

\paragraph{Ultra-large-scale phase retrieval.} In ultra-large-scale imaging applications such as 4K (4096$\times$2160 pixels) or 8K (7680$\times$4320 pixels), most reconstruction algorithms are not applicable due to either highly large memory requirement or extremely long running time. Nevertheless, the reported \emph{LPR} technique still works well in such applications. As a demonstration, we implemented a simulation of 8K-level CDP (5 modulations), using an 8K outer space color image as ground truth (released by NASA using the Hubble Telescope). Its spatial resolution is 7680$\times$4320 (each color channel) with in total 33.1 million pixels. Figure \ref{fig:ultra} presents the reconstruction results of AP and \emph{LPR}, with the input SNR being 5dB. The colse-ups show that the result of AP is drowned out by measurement noise, leading to dimness and loss of target details. In comparison, \emph{LPR} outperforms a lot with strong robustness. Both their running time lie in the minute level. Another set of 8K reconstruction results is shown in Fig. S8 (supplementary information).

\section*{Methods} 
Following optimization theory, the phase retrieval task can be modeled as
\begin{equation}
\label{eq:5}
\hat{u}=\arg \min _{u} f(u)+\lambda g(u),
\end{equation}
where $u$ denotes the target complex filed to be recovered, $f(u)$ is a data-fidelity regularizer that ensures consistency between the reconstructed result and measurements, and $g(u)$ is a regularizer that imposes certain statistical prior knowledge. Conventionally, Eq. (\ref{eq:5}) is solved following the first-order proximal gradient methods, such as ISTA and ADMM that are time-consuming to calculate gradients in large-scale nonlinear tasks \cite{liu2018rank}. In this work, instead, we employ the efficient generalized-alternating-projection (GAP) strategy \cite{liu2018rank} to transform Eq. (\ref{eq:5}) with fewer variables to
\begin{equation}
\label{eq:6}
\begin{array}{c}
(u, v)=\operatorname{argmin} 1 / 2\|u-v\|_{2}^{2}+\lambda g(v) \\
\text { s.t. } I=|A u|^{2},
\end{array}
\end{equation}
where $v$ is an auxiliary variable balancing the data fidelity term and prior regularization, $A$ denotes measurement matrix, and $I$ represents measurement. The difference between the conventional ADMM and GAP optimization is the constraint on the measurement\cite{liu2018rank}. ADMM minimizes $
\left\|I-|A u|^{2}\right\|$, while GAP imposes the constraint $I=|A u|^{2}$.

To tackle the large-scale phase retrieval task, we extend the efficient plug-and-play (PNP) optimization framework \cite{venkatakrishnan2013plug} from real space to nonlinear complex space. Fundamentally, PNP decomposes optimization into two separate sub-problems including measurement formation and prior regularization, so as to incorporating inverse recovery solvers together with various image enhancing solvers to improve reconstruction accuracy, providing high flexibility for different applications.
Mathematically, Eq. (\ref{eq:6}) is decomposed into the following two sub-problems, to alternatively update the two variables $u$ and $v$.

\begin{itemize}

\item Updating $u$: given $v^{(k)}$, $u^{(k+1)}$ is updated via a Euclidean projection of $v^{(k)}$ on the manifold $I=|A u|^{2}$ as
\begin{equation}
\label{eq:7}
u^{k+1}=v^{(k)}+PR\left(I-|A v|^{2}\right),
\end{equation}
where $PR$ is phase retrieval solver. Considering its great generalization ability on various imaging modalities and low computational complexity, we employ the AP method as the $PR$ solver. It alternates between the target and observation planes allowing to incorporate any information available for the variables, providing low sampling rate requirement.

\item Updating $v$: given $u^{(k+1)}$, $v^{(k+1)}$ is updated by an image enhancing solver $EN$ as
\begin{equation}
\label{eq:8}
v^{k+1}=E N\left(u^{k+1}\right).
\end{equation}
Although the iterative image enhancing research has made great progress in recent years with such as non-local optimization and dictionary learning \cite{elad2006image}, they maintain high computational complexity for large-scale reconstruction \cite{zhang2017beyond}. In this work, considering its flexible and fast solution, we employed the deep learning based FFDNET \cite{zhang2018ffdnet} to tackle the sub-problem with high fidelity and self-adaptation. The neural network consists of a series of 3$\times$3 convolution layers. Each layer is composed of a specific combination of three types of operations including convolution, rectified linear units and batch normalization. The architecture provides a balanced tradeoff between noise suppression and detail fidelity. While an image is input into the network, it is first down sampled into several sub-blocks, which then flow through the network for quality enhancement. Finally, these optimized blocks are stitched together to the original size. Such a workflow enables its great generalization ability on different image size.

\end{itemize}

After initialization, the variables are updated alternatively following Eq. (\ref{eq:7}) and Eq. (\ref{eq:8}). When the intensity difference of reconstructed image between two successive iterations is smaller than a given threshold, the iteration stops with convergence. Since both the two solvers $PR$ and $EN$ are highly efficient and flexible, the entire reconstruction maintains low computational complexity and great generalization. The demo code has been released at \url{bianlab.github.io}.

\section*{Conclusion and Discussion}
In this work, we engaged to tackle the large-scale phase retrieval problem, and reported a generalized \emph{LPR} optimization technique with low computational complexity and strong robustness. It extends the efficient PNP-GAP framework from real space to nonlinear complex space, and incorporates the alternating projection solver and enhancing neural network. As validated by extensive simulations and experiments on three different computational phase imaging modalities (CDI, CDP and FPM), \emph{LPR} exhibits unique advantages in large-scale phase retrieval tasks with high fidelity and efficiency.

The \emph{LPR} technique can be further extended. First, it involves multiple algorithm parameters that are currently adjusted manually. We can introduce the reinforcement learning technique \cite{wei2020tuning} in our future work to automatically adjust these parameters for best performance. Second, \emph{LPR} is sensitive to initialization, especially under low sampling rate. The optimal spectral initialization\cite{luo2019optimal} technique can be incorporated for stronger robustness. Third, it is interesting to investigate the influence of employing other image enhancing solvers such as superresolution neural network \cite{wang2020deep} and deblurring network. This may open new insights for phase retrieval with boosted quality.

\vspace{5mm}

\bibliographystyle{naturemag}
\bibliography{references}

\begin{thebibliography}{10}
\expandafter\ifx\csname url\endcsname\relax
  \def\url#1{\texttt{#1}}\fi
\expandafter\ifx\csname urlprefix\endcsname\relax\def\urlprefix{URL }\fi
\providecommand{\bibinfo}[2]{#2}
\providecommand{\eprint}[2][]{\url{#2}}

\bibitem{pahlevaninezhad2018nano}
\bibinfo{author}{Pahlevaninezhad, H.} \emph{et~al.}
\newblock \bibinfo{title}{Nano-optic endoscope for high-resolution optical
  coherence tomography in vivo}.
\newblock \emph{\bibinfo{journal}{Nat. Photonics}}
  \textbf{\bibinfo{volume}{12}}, \bibinfo{pages}{540--547}
  (\bibinfo{year}{2018}).

\bibitem{lombardini2018high}
\bibinfo{author}{Lombardini, A.} \emph{et~al.}
\newblock \bibinfo{title}{High-resolution multimodal flexible coherent {R}aman
  endoscope}.
\newblock \emph{\bibinfo{journal}{Light: Sci. Appl.}}
  \textbf{\bibinfo{volume}{7}}, \bibinfo{pages}{1--8} (\bibinfo{year}{2018}).

\bibitem{zheng2013wide}
\bibinfo{author}{Zheng, G.}, \bibinfo{author}{Horstmeyer, R.} \&
  \bibinfo{author}{Yang, C.}
\newblock \bibinfo{title}{Wide-field, high-resolution {F}ourier ptychographic
  microscopy}.
\newblock \emph{\bibinfo{journal}{Nat. Photonics}}
  \textbf{\bibinfo{volume}{7}}, \bibinfo{pages}{739--745}
  (\bibinfo{year}{2013}).

\bibitem{fan2019video}
\bibinfo{author}{Fan, J.} \emph{et~al.}
\newblock \bibinfo{title}{Video-rate imaging of biological dynamics at
  centimetre scale and micrometre resolution}.
\newblock \emph{\bibinfo{journal}{Nat. Photonics}}
  \textbf{\bibinfo{volume}{13}}, \bibinfo{pages}{809--816}
  (\bibinfo{year}{2019}).

\bibitem{wang2011space}
\bibinfo{author}{Wang, W.-Q.}
\newblock \bibinfo{title}{Space--time coding {MIMO-OFDM} {SAR} for
  high-resolution imaging}.
\newblock \emph{\bibinfo{journal}{IEEE T. Geosci. Remote}}
  \textbf{\bibinfo{volume}{49}}, \bibinfo{pages}{3094--3104}
  (\bibinfo{year}{2011}).

\bibitem{brady2012multiscale}
\bibinfo{author}{Brady, D.~J.} \emph{et~al.}
\newblock \bibinfo{title}{Multiscale gigapixel photography}.
\newblock \emph{\bibinfo{journal}{Nature}} \textbf{\bibinfo{volume}{486}},
  \bibinfo{pages}{386--389} (\bibinfo{year}{2012}).

\bibitem{wang2016computational}
\bibinfo{author}{Wang, H.} \emph{et~al.}
\newblock \bibinfo{title}{Computational out-of-focus imaging increases the
  space--bandwidth product in lens-based coherent microscopy}.
\newblock \emph{\bibinfo{journal}{Optica}} \textbf{\bibinfo{volume}{3}},
  \bibinfo{pages}{1422--1429} (\bibinfo{year}{2016}).

\bibitem{yuan2020plug}
\bibinfo{author}{Yuan, X.}, \bibinfo{author}{Liu, Y.}, \bibinfo{author}{Suo,
  J.} \& \bibinfo{author}{Dai, Q.}
\newblock \bibinfo{title}{Plug-and-play algorithms for large-scale snapshot
  compressive imaging}.
\newblock In \emph{\bibinfo{booktitle}{Conference on Computer Vision and
  Pattern Recognition (CVPR)}}, \bibinfo{pages}{1447--1457}
  (\bibinfo{year}{2020}).

\bibitem{shechtman2015phase}
\bibinfo{author}{Shechtman, Y.} \emph{et~al.}
\newblock \bibinfo{title}{Phase retrieval with application to optical imaging:
  a contemporary overview}.
\newblock \emph{\bibinfo{journal}{IEEE Signal Proc. Mag.}}
  \textbf{\bibinfo{volume}{32}}, \bibinfo{pages}{87--109}
  (\bibinfo{year}{2015}).

\bibitem{miao1999extending}
\bibinfo{author}{Miao, J.}, \bibinfo{author}{Charalambous, P.},
  \bibinfo{author}{Kirz, J.} \& \bibinfo{author}{Sayre, D.}
\newblock \bibinfo{title}{Extending the methodology of {X}-ray crystallography
  to allow imaging of micrometre-sized non-crystalline specimens}.
\newblock \emph{\bibinfo{journal}{Nature}} \textbf{\bibinfo{volume}{400}},
  \bibinfo{pages}{342--344} (\bibinfo{year}{1999}).

\bibitem{candes2015phase}
\bibinfo{author}{Candes, E.~J.}, \bibinfo{author}{Li, X.} \&
  \bibinfo{author}{Soltanolkotabi, M.}
\newblock \bibinfo{title}{Phase retrieval from coded diffraction patterns}.
\newblock \emph{\bibinfo{journal}{Appl. Comput. Harmon. A.}}
  \textbf{\bibinfo{volume}{39}}, \bibinfo{pages}{277--299}
  (\bibinfo{year}{2015}).

\bibitem{katz2014non}
\bibinfo{author}{Katz, O.}, \bibinfo{author}{Heidmann, P.},
  \bibinfo{author}{Fink, M.} \& \bibinfo{author}{Gigan, S.}
\newblock \bibinfo{title}{Non-invasive single-shot imaging through scattering
  layers and around corners via speckle correlations}.
\newblock \emph{\bibinfo{journal}{Nat. Photonics}}
  \textbf{\bibinfo{volume}{8}}, \bibinfo{pages}{784--790}
  (\bibinfo{year}{2014}).

\bibitem{gerchberg1972practical}
\bibinfo{author}{Gerchberg, R.~W.}
\newblock \bibinfo{title}{A practical algorithm for the determination of phase
  from image and diffraction plane pictures}.
\newblock \emph{\bibinfo{journal}{Optik}} \textbf{\bibinfo{volume}{35}},
  \bibinfo{pages}{237--246} (\bibinfo{year}{1972}).

\bibitem{fienup1982phase}
\bibinfo{author}{Fienup, J.~R.}
\newblock \bibinfo{title}{Phase retrieval algorithms: a comparison}.
\newblock \emph{\bibinfo{journal}{Appl. Optics}} \textbf{\bibinfo{volume}{21}},
  \bibinfo{pages}{2758--2769} (\bibinfo{year}{1982}).

\bibitem{candes2013phaselift}
\bibinfo{author}{Candes, E.~J.}, \bibinfo{author}{Strohmer, T.} \&
  \bibinfo{author}{Voroninski, V.}
\newblock \bibinfo{title}{Phaselift: Exact and stable signal recovery from
  magnitude measurements via convex programming}.
\newblock \emph{\bibinfo{journal}{Commun. Pur. Appl. Math.}}
  \textbf{\bibinfo{volume}{66}}, \bibinfo{pages}{1241--1274}
  (\bibinfo{year}{2013}).

\bibitem{vandenberghe1996semidefinite}
\bibinfo{author}{Vandenberghe, L.} \& \bibinfo{author}{Boyd, S.}
\newblock \bibinfo{title}{Semidefinite programming}.
\newblock \emph{\bibinfo{journal}{SIAM Rev.}} \textbf{\bibinfo{volume}{38}},
  \bibinfo{pages}{49--95} (\bibinfo{year}{1996}).

\bibitem{candes2015phasewf}
\bibinfo{author}{Candes, E.~J.}, \bibinfo{author}{Li, X.} \&
  \bibinfo{author}{Soltanolkotabi, M.}
\newblock \bibinfo{title}{Phase retrieval via {W}irtinger flow: Theory and
  algorithms}.
\newblock \emph{\bibinfo{journal}{IEEE T. Inform. Theory}}
  \textbf{\bibinfo{volume}{61}}, \bibinfo{pages}{1985--2007}
  (\bibinfo{year}{2015}).

\bibitem{chen2015solving}
\bibinfo{author}{Chen, Y.} \& \bibinfo{author}{Candes, E.}
\newblock \bibinfo{title}{Solving random quadratic systems of equations is
  nearly as easy as solving linear systems}.
\newblock In \emph{\bibinfo{booktitle}{International Conference on Neural
  Information Processing Systems (NIPS)}}, \bibinfo{pages}{739--747}
  (\bibinfo{year}{2015}).

\bibitem{zeng2020coordinate}
\bibinfo{author}{Zeng, W.-J.} \& \bibinfo{author}{So, H.-C.}
\newblock \bibinfo{title}{Coordinate descent algorithms for phase retrieval}.
\newblock \emph{\bibinfo{journal}{Signal Process.}}
  \textbf{\bibinfo{volume}{169}}, \bibinfo{pages}{107418}
  (\bibinfo{year}{2020}).

\bibitem{katkovnik2017phase}
\bibinfo{author}{Katkovnik, V.}
\newblock \bibinfo{title}{Phase retrieval from noisy data based on sparse
  approximation of object phase and amplitude}.
\newblock \emph{\bibinfo{journal}{arXiv preprint arXiv:1709.01071}}
  (\bibinfo{year}{2017}).

\bibitem{metzler2016bm3d}
\bibinfo{author}{Metzler, C.~A.}, \bibinfo{author}{Maleki, A.} \&
  \bibinfo{author}{Baraniuk, R.~G.}
\newblock \bibinfo{title}{{BM3D-PRGAMP}: Compressive phase retrieval based on
  {BM3D} denoising}.
\newblock In \emph{\bibinfo{booktitle}{International Conference on Image
  Processing (ICIP)}}, \bibinfo{pages}{2504--2508}
  (\bibinfo{organization}{IEEE}, \bibinfo{year}{2016}).

\bibitem{metzler2018prdeep}
\bibinfo{author}{Metzler, C.}, \bibinfo{author}{Schniter, P.},
  \bibinfo{author}{Veeraraghavan, A.} \emph{et~al.}
\newblock \bibinfo{title}{pr{D}eep: robust phase retrieval with a flexible deep
  network}.
\newblock In \emph{\bibinfo{booktitle}{International Conference on Machine
  Learning (ICML)}}, \bibinfo{pages}{3501--3510} (\bibinfo{organization}{PMLR},
  \bibinfo{year}{2018}).

\bibitem{rivenson2018phase}
\bibinfo{author}{Rivenson, Y.}, \bibinfo{author}{Zhang, Y.},
  \bibinfo{author}{G{\"u}nayd{\i}n, H.}, \bibinfo{author}{Teng, D.} \&
  \bibinfo{author}{Ozcan, A.}
\newblock \bibinfo{title}{Phase recovery and holographic image reconstruction
  using deep learning in neural networks}.
\newblock \emph{\bibinfo{journal}{Light: Sci. Appl.}}
  \textbf{\bibinfo{volume}{7}}, \bibinfo{pages}{17141--17141}
  (\bibinfo{year}{2018}).

\bibitem{kappeler2017ptychnet}
\bibinfo{author}{Kappeler, A.}, \bibinfo{author}{Ghosh, S.},
  \bibinfo{author}{Holloway, J.}, \bibinfo{author}{Cossairt, O.} \&
  \bibinfo{author}{Katsaggelos, A.}
\newblock \bibinfo{title}{Ptychnet: {CNN} based {F}ourier ptychography}.
\newblock In \emph{\bibinfo{booktitle}{International Conference on Image
  Processing (ICIP)}}, \bibinfo{pages}{1712--1716}
  (\bibinfo{organization}{IEEE}, \bibinfo{year}{2017}).

\bibitem{venkatakrishnan2013plug}
\bibinfo{author}{Venkatakrishnan, S.~V.}, \bibinfo{author}{Bouman, C.~A.} \&
  \bibinfo{author}{Wohlberg, B.}
\newblock \bibinfo{title}{Plug-and-play priors for model based reconstruction}.
\newblock In \emph{\bibinfo{booktitle}{Global Conference on Signal and
  Information Processing (GlobalSIP)}}, \bibinfo{pages}{945--948}
  (\bibinfo{organization}{IEEE}, \bibinfo{year}{2013}).

\bibitem{liao2014generalized}
\bibinfo{author}{Liao, X.}, \bibinfo{author}{Li, H.} \& \bibinfo{author}{Carin,
  L.}
\newblock \bibinfo{title}{Generalized alternating projection for weighted-2,1
  minimization with applications to model-based compressive sensing}.
\newblock \emph{\bibinfo{journal}{SIAM J. Imaging Sci.}}
  \textbf{\bibinfo{volume}{7}}, \bibinfo{pages}{797--823}
  (\bibinfo{year}{2014}).

\bibitem{yuan2016generalized}
\bibinfo{author}{Yuan, X.}
\newblock \bibinfo{title}{Generalized alternating projection based total
  variation minimization for compressive sensing}.
\newblock In \emph{\bibinfo{booktitle}{International Conference on Image
  Processing (ICIP)}}, \bibinfo{pages}{2539--2543}
  (\bibinfo{organization}{IEEE}, \bibinfo{year}{2016}).

\bibitem{bioucas2007new}
\bibinfo{author}{Bioucas-Dias, J.~M.} \& \bibinfo{author}{Figueiredo, M.~A.}
\newblock \bibinfo{title}{A new {TwIST}: Two-step iterative
  shrinkage/thresholding algorithms for image restoration}.
\newblock \emph{\bibinfo{journal}{IEEE T. Image Process.}}
  \textbf{\bibinfo{volume}{16}}, \bibinfo{pages}{2992--3004}
  (\bibinfo{year}{2007}).

\bibitem{liu2018rank}
\bibinfo{author}{Liu, Y.}, \bibinfo{author}{Yuan, X.}, \bibinfo{author}{Suo,
  J.}, \bibinfo{author}{Brady, D.~J.} \& \bibinfo{author}{Dai, Q.}
\newblock \bibinfo{title}{Rank minimization for snapshot compressive imaging}.
\newblock \emph{\bibinfo{journal}{IEEE T. Pattern Anal.}}
  \textbf{\bibinfo{volume}{41}}, \bibinfo{pages}{2990--3006}
  (\bibinfo{year}{2018}).

\bibitem{goldstein2018phasemax}
\bibinfo{author}{Goldstein, T.} \& \bibinfo{author}{Studer, C.}
\newblock \bibinfo{title}{Phasemax: Convex phase retrieval via basis pursuit}.
\newblock \emph{\bibinfo{journal}{IEEE T. Inform. Theory}}
  \textbf{\bibinfo{volume}{64}}, \bibinfo{pages}{2675--2689}
  (\bibinfo{year}{2018}).

\bibitem{dhifallah2017phase}
\bibinfo{author}{Dhifallah, O.}, \bibinfo{author}{Thrampoulidis, C.} \&
  \bibinfo{author}{Lu, Y.~M.}
\newblock \bibinfo{title}{Phase retrieval via linear programming: Fundamental
  limits and algorithmic improvements}.
\newblock In \emph{\bibinfo{booktitle}{Annual Allerton Conference on
  Communication, Control, and Computing (Allerton)}},
  \bibinfo{pages}{1071--1077} (\bibinfo{organization}{IEEE},
  \bibinfo{year}{2017}).

\bibitem{yuan2017phase}
\bibinfo{author}{Yuan, Z.} \& \bibinfo{author}{Wang, H.}
\newblock \bibinfo{title}{Phase retrieval via reweighted {W}irtinger flow}.
\newblock \emph{\bibinfo{journal}{Appl. Optics}} \textbf{\bibinfo{volume}{56}},
  \bibinfo{pages}{2418--2427} (\bibinfo{year}{2017}).

\bibitem{wang2017solving}
\bibinfo{author}{Wang, G.}, \bibinfo{author}{Giannakis, G.~B.} \&
  \bibinfo{author}{Eldar, Y.~C.}
\newblock \bibinfo{title}{Solving systems of random quadratic equations via
  truncated amplitude flow}.
\newblock \emph{\bibinfo{journal}{IEEE T. Inform. Theory}}
  \textbf{\bibinfo{volume}{64}}, \bibinfo{pages}{773--794}
  (\bibinfo{year}{2017}).

\bibitem{wang2018phase}
\bibinfo{author}{Wang, G.}, \bibinfo{author}{Giannakis, G.~B.},
  \bibinfo{author}{Saad, Y.} \& \bibinfo{author}{Chen, J.}
\newblock \bibinfo{title}{Phase retrieval via reweighted amplitude flow}.
\newblock \emph{\bibinfo{journal}{IEEE T. Signal Proces.}}
  \textbf{\bibinfo{volume}{66}}, \bibinfo{pages}{2818--2833}
  (\bibinfo{year}{2018}).

\bibitem{zeng2017coordinate}
\bibinfo{author}{Zeng, W.-J.} \& \bibinfo{author}{So, H.-C.}
\newblock \bibinfo{title}{Coordinate descent algorithms for phase retrieval}.
\newblock \emph{\bibinfo{journal}{arXiv preprint arXiv:1706.03474}}
  (\bibinfo{year}{2017}).

\bibitem{wei2015solving}
\bibinfo{author}{Wei, K.}
\newblock \bibinfo{title}{Solving systems of phaseless equations via {K}aczmarz
  methods: A proof of concept study}.
\newblock \emph{\bibinfo{journal}{Inverse Probl.}}
  \textbf{\bibinfo{volume}{31}}, \bibinfo{pages}{125008}
  (\bibinfo{year}{2015}).

\bibitem{chandra2019phasepack}
\bibinfo{author}{Chandra, R.}, \bibinfo{author}{Goldstein, T.} \&
  \bibinfo{author}{Studer, C.}
\newblock \bibinfo{title}{Phasepack: A phase retrieval library}.
\newblock In \emph{\bibinfo{booktitle}{International conference on Sampling
  Theory and Applications (SampTA)}}, \bibinfo{pages}{1--5}
  (\bibinfo{organization}{IEEE}, \bibinfo{year}{2019}).

\bibitem{wang2004image}
\bibinfo{author}{Wang, Z.}, \bibinfo{author}{Bovik, A.~C.},
  \bibinfo{author}{Sheikh, H.~R.} \& \bibinfo{author}{Simoncelli, E.~P.}
\newblock \bibinfo{title}{Image quality assessment: from error visibility to
  structural similarity}.
\newblock \emph{\bibinfo{journal}{IEEE T. Image Process.}}
  \textbf{\bibinfo{volume}{13}}, \bibinfo{pages}{600--612}
  (\bibinfo{year}{2004}).

\bibitem{agustsson2017ntire}
\bibinfo{author}{Agustsson, E.} \& \bibinfo{author}{Timofte, R.}
\newblock \bibinfo{title}{Ntire 2017 challenge on single image
  super-resolution: Dataset and study}.
\newblock In \emph{\bibinfo{booktitle}{Conference on Computer Vision and
  Pattern Recognition (CVPR)}}, \bibinfo{pages}{126--135}
  (\bibinfo{year}{2017}).

\bibitem{miao2015beyond}
\bibinfo{author}{Miao, J.}, \bibinfo{author}{Ishikawa, T.},
  \bibinfo{author}{Robinson, I.~K.} \& \bibinfo{author}{Murnane, M.~M.}
\newblock \bibinfo{title}{Beyond crystallography: Diffractive imaging using
  coherent {X}-ray light sources}.
\newblock \emph{\bibinfo{journal}{Science}} \textbf{\bibinfo{volume}{348}},
  \bibinfo{pages}{530--535} (\bibinfo{year}{2015}).

\bibitem{lo2018situ}
\bibinfo{author}{Lo, Y.~H.} \emph{et~al.}
\newblock \bibinfo{title}{In situ coherent diffractive imaging}.
\newblock \emph{\bibinfo{journal}{Nat. Commun.}} \textbf{\bibinfo{volume}{9}},
  \bibinfo{pages}{1--10} (\bibinfo{year}{2018}).

\bibitem{blood_cell}
\bibinfo{author}{Choksawatdikorn}.
\newblock \bibinfo{title}{Blood cells under microscope view for histology
  education}.
\newblock
  \bibinfo{howpublished}{\url{https://www.shutterstock.com/zh/image-photo/blood-cells-under-microscope-view-histology-1102617128}}
  (\bibinfo{year}{2020}).
\newblock \bibinfo{note}{[Online; accessed 5-November-2020]}.

\bibitem{bian2015fourier}
\bibinfo{author}{Bian, L.} \emph{et~al.}
\newblock \bibinfo{title}{Fourier ptychographic reconstruction using
  {W}irtinger flow optimization}.
\newblock \emph{\bibinfo{journal}{Opt. Express}} \textbf{\bibinfo{volume}{23}},
  \bibinfo{pages}{4856--4866} (\bibinfo{year}{2015}).

\bibitem{elad2006image}
\bibinfo{author}{Elad, M.} \& \bibinfo{author}{Aharon, M.}
\newblock \bibinfo{title}{Image denoising via sparse and redundant
  representations over learned dictionaries}.
\newblock \emph{\bibinfo{journal}{IEEE T. Image Process.}}
  \textbf{\bibinfo{volume}{15}}, \bibinfo{pages}{3736--3745}
  (\bibinfo{year}{2006}).

\bibitem{zhang2017beyond}
\bibinfo{author}{Zhang, K.}, \bibinfo{author}{Zuo, W.}, \bibinfo{author}{Chen,
  Y.}, \bibinfo{author}{Meng, D.} \& \bibinfo{author}{Zhang, L.}
\newblock \bibinfo{title}{Beyond a gaussian denoiser: Residual learning of deep
  {CNN} for image denoising}.
\newblock \emph{\bibinfo{journal}{IEEE T. Image Process.}}
  \textbf{\bibinfo{volume}{26}}, \bibinfo{pages}{3142--3155}
  (\bibinfo{year}{2017}).

\bibitem{zhang2018ffdnet}
\bibinfo{author}{Zhang, K.}, \bibinfo{author}{Zuo, W.} \&
  \bibinfo{author}{Zhang, L.}
\newblock \bibinfo{title}{{FFDNet}: Toward a fast and flexible solution for
  {CNN}-based image denoising}.
\newblock \emph{\bibinfo{journal}{IEEE T. Image Process.}}
  \textbf{\bibinfo{volume}{27}}, \bibinfo{pages}{4608--4622}
  (\bibinfo{year}{2018}).

\bibitem{wei2020tuning}
\bibinfo{author}{Wei, K.} \emph{et~al.}
\newblock \bibinfo{title}{Tuning-free plug-and-play proximal algorithm for
  inverse imaging problems}.
\newblock In \emph{\bibinfo{booktitle}{International Conference on Machine
  Learning (ICML)}}, \bibinfo{pages}{10158--10169}
  (\bibinfo{organization}{PMLR}, \bibinfo{year}{2020}).

\bibitem{luo2019optimal}
\bibinfo{author}{Luo, W.}, \bibinfo{author}{Alghamdi, W.} \&
  \bibinfo{author}{Lu, Y.~M.}
\newblock \bibinfo{title}{Optimal spectral initialization for signal recovery
  with applications to phase retrieval}.
\newblock \emph{\bibinfo{journal}{IEEE T. Signal Proces.}}
  \textbf{\bibinfo{volume}{67}}, \bibinfo{pages}{2347--2356}
  (\bibinfo{year}{2019}).

\bibitem{wang2020deep}
\bibinfo{author}{Wang, Z.}, \bibinfo{author}{Chen, J.} \& \bibinfo{author}{Hoi,
  S.~C.}
\newblock \bibinfo{title}{Deep learning for image super-resolution: A survey}.
\newblock \emph{\bibinfo{journal}{IEEE T. Pattern Anal.}}
  (\bibinfo{year}{2020}).

\end{thebibliography}


\begin{thebibliography}{10}
\providecommand{\url}[1]{#1}
\csname url@samestyle\endcsname
\providecommand{\newblock}{\relax}
\providecommand{\bibinfo}[2]{#2}
\providecommand{\BIBentrySTDinterwordspacing}{\spaceskip=0pt\relax}
\providecommand{\BIBentryALTinterwordstretchfactor}{4}
\providecommand{\BIBentryALTinterwordspacing}{\spaceskip=\fontdimen2\font plus
\BIBentryALTinterwordstretchfactor\fontdimen3\font minus
  \fontdimen4\font\relax}
\providecommand{\BIBforeignlanguage}[2]{{%
\expandafter\ifx\csname l@#1\endcsname\relax
\typeout{** WARNING: IEEEtran.bst: No hyphenation pattern has been}%
\typeout{** loaded for the language `#1'. Using the pattern for}%
\typeout{** the default language instead.}%
\else
\language=\csname l@#1\endcsname
\fi
#2}}
\providecommand{\BIBdecl}{\relax}
\BIBdecl

\bibitem{fan2019video}
J.~Fan, J.~Suo, J.~Wu, H.~Xie, Y.~Shen, F.~Chen, G.~Wang, L.~Cao, G.~Jin, Q.~He
  \emph{et~al.}, ``Video-rate imaging of biological dynamics at centimetre
  scale and micrometre resolution,'' \emph{Nature Photonics}, vol.~13, no.~11,
  pp. 809--816, 2019.

\bibitem{brady2012multiscale}
D.~J. Brady, M.~E. Gehm, R.~A. Stack, D.~L. Marks, D.~S. Kittle, D.~R. Golish,
  E.~Vera, and S.~D. Feller, ``Multiscale gigapixel photography,''
  \emph{Nature}, vol. 486, no. 7403, pp. 386--389, 2012.

\bibitem{zheng2013wide}
G.~Zheng, R.~Horstmeyer, and C.~Yang, ``Wide-field, high-resolution fourier
  ptychographic microscopy,'' \emph{Nature photonics}, vol.~7, no.~9, pp.
  739--745, 2013.

\bibitem{wang2016computational}
H.~Wang, Z.~G{\"o}r{\"o}cs, W.~Luo, Y.~Zhang, Y.~Rivenson, L.~A. Bentolila, and
  A.~Ozcan, ``Computational out-of-focus imaging increases the space--bandwidth
  product in lens-based coherent microscopy,'' \emph{Optica}, vol.~3, no.~12,
  pp. 1422--1429, 2016.

\bibitem{yuan2020plug}
X.~Yuan, Y.~Liu, J.~Suo, and Q.~Dai, ``Plug-and-play algorithms for large-scale
  snapshot compressive imaging,'' in \emph{Proceedings of the IEEE/CVF
  Conference on Computer Vision and Pattern Recognition}, 2020, pp. 1447--1457.

\bibitem{miao1999extending}
J.~Miao, P.~Charalambous, J.~Kirz, and D.~Sayre, ``Extending the methodology of
  x-ray crystallography to allow imaging of micrometre-sized non-crystalline
  specimens,'' \emph{Nature}, vol. 400, no. 6742, pp. 342--344, 1999.

\bibitem{candes2015phase}
E.~J. Candes, X.~Li, and M.~Soltanolkotabi, ``Phase retrieval from coded
  diffraction patterns,'' \emph{Applied and Computational Harmonic Analysis},
  vol.~39, no.~2, pp. 277--299, 2015.

\bibitem{katz2014non}
O.~Katz, P.~Heidmann, M.~Fink, and S.~Gigan, ``Non-invasive single-shot imaging
  through scattering layers and around corners via speckle correlations,''
  \emph{Nature photonics}, vol.~8, no.~10, pp. 784--790, 2014.

\bibitem{rajaei2016intensity}
B.~Rajaei, E.~W. Tramel, S.~Gigan, F.~Krzakala, and L.~Daudet, ``Intensity-only
  optical compressive imaging using a multiply scattering material and a double
  phase retrieval approach,'' in \emph{2016 IEEE International Conference on
  Acoustics, Speech and Signal Processing (ICASSP)}.\hskip 1em plus 0.5em minus
  0.4em\relax IEEE, 2016, pp. 4054--4058.

\bibitem{fienup1982phase}
J.~R. Fienup, ``Phase retrieval algorithms: a comparison,'' \emph{Applied
  optics}, vol.~21, no.~15, pp. 2758--2769, 1982.

\bibitem{candes2013phaselift}
E.~J. Candes, T.~Strohmer, and V.~Voroninski, ``Phaselift: Exact and stable
  signal recovery from magnitude measurements via convex programming,''
  \emph{Communications on Pure and Applied Mathematics}, vol.~66, no.~8, pp.
  1241--1274, 2013.

\bibitem{vandenberghe1996semidefinite}
L.~Vandenberghe and S.~Boyd, ``Semidefinite programming,'' \emph{SIAM review},
  vol.~38, no.~1, pp. 49--95, 1996.

\bibitem{candes2015phasewf}
E.~J. Candes, X.~Li, and M.~Soltanolkotabi, ``Phase retrieval via wirtinger
  flow: Theory and algorithms,'' \emph{IEEE Transactions on Information
  Theory}, vol.~61, no.~4, pp. 1985--2007, 2015.

\bibitem{chen2015solving}
Y.~Chen and E.~Candes, ``Solving random quadratic systems of equations is
  nearly as easy as solving linear systems,'' in \emph{Advances in Neural
  Information Processing Systems}, 2015, pp. 739--747.

\bibitem{zeng2020coordinate}
W.-J. Zeng and H.-C. So, ``Coordinate descent algorithms for phase retrieval,''
  \emph{Signal Processing}, vol. 169, p. 107418, 2020.

\bibitem{mukherjee2012iterative}
S.~Mukherjee and C.~S. Seelamantula, ``An iterative algorithm for phase
  retrieval with sparsity constraints: application to frequency domain optical
  coherence tomography,'' in \emph{2012 IEEE International Conference on
  Acoustics, Speech and Signal Processing (ICASSP)}.\hskip 1em plus 0.5em minus
  0.4em\relax IEEE, 2012, pp. 553--556.

\bibitem{katkovnik2017phase}
V.~Katkovnik, ``Phase retrieval from noisy data based on sparse approximation
  of object phase and amplitude,'' \emph{arXiv preprint arXiv:1709.01071},
  2017.

\bibitem{metzler2016bm3d}
C.~A. Metzler, A.~Maleki, and R.~G. Baraniuk, ``Bm3d-prgamp: Compressive phase
  retrieval based on bm3d denoising,'' in \emph{2016 IEEE International
  Conference on Image Processing (ICIP)}.\hskip 1em plus 0.5em minus
  0.4em\relax IEEE, 2016, pp. 2504--2508.

\bibitem{metzler2018prdeep}
C.~A. Metzler, P.~Schniter, A.~Veeraraghavan, and R.~G. Baraniuk, ``prdeep:
  Robust phase retrieval with a flexible deep network,'' \emph{arXiv preprint
  arXiv:1803.00212}, 2018.

\bibitem{goy2018low}
A.~Goy, K.~Arthur, S.~Li, and G.~Barbastathis, ``Low photon count phase
  retrieval using deep learning,'' \emph{Physical review letters}, vol. 121,
  no.~24, p. 243902, 2018.

\bibitem{rivenson2018phase}
Y.~Rivenson, Y.~Zhang, H.~G{\"u}nayd{\i}n, D.~Teng, and A.~Ozcan, ``Phase
  recovery and holographic image reconstruction using deep learning in neural
  networks,'' \emph{Light: Science \& Applications}, vol.~7, no.~2, pp.
  17\,141--17\,141, 2018.

\bibitem{kappeler2017ptychnet}
A.~Kappeler, S.~Ghosh, J.~Holloway, O.~Cossairt, and A.~Katsaggelos,
  ``Ptychnet: Cnn based fourier ptychography,'' in \emph{2017 IEEE
  International Conference on Image Processing (ICIP)}.\hskip 1em plus 0.5em
  minus 0.4em\relax IEEE, 2017, pp. 1712--1716.

\bibitem{liao2014generalized}
X.~Liao, H.~Li, and L.~Carin, ``Generalized alternating projection for
  weighted-2,1 minimization with applications to model-based compressive
  sensing,'' \emph{SIAM Journal on Imaging Sciences}, vol.~7, no.~2, pp.
  797--823, 2014.

\bibitem{yuan2016generalized}
X.~Yuan, ``Generalized alternating projection based total variation
  minimization for compressive sensing,'' in \emph{2016 IEEE International
  Conference on Image Processing (ICIP)}.\hskip 1em plus 0.5em minus
  0.4em\relax IEEE, 2016, pp. 2539--2543.

\bibitem{bioucas2007new}
J.~M. Bioucas-Dias and M.~A. Figueiredo, ``A new twist: Two-step iterative
  shrinkage/thresholding algorithms for image restoration,'' \emph{IEEE
  Transactions on Image processing}, vol.~16, no.~12, pp. 2992--3004, 2007.

\bibitem{beck2009fast}
A.~Beck and M.~Teboulle, ``A fast iterative shrinkage-thresholding algorithm
  for linear inverse problems,'' \emph{SIAM journal on imaging sciences},
  vol.~2, no.~1, pp. 183--202, 2009.

\bibitem{boyd2011distributed}
S.~Boyd, N.~Parikh, and E.~Chu, \emph{Distributed optimization and statistical
  learning via the alternating direction method of multipliers}.\hskip 1em plus
  0.5em minus 0.4em\relax Now Publishers Inc, 2011.

\bibitem{venkatakrishnan2013plug}
S.~V. Venkatakrishnan, C.~A. Bouman, and B.~Wohlberg, ``Plug-and-play priors
  for model based reconstruction,'' in \emph{2013 IEEE Global Conference on
  Signal and Information Processing}.\hskip 1em plus 0.5em minus 0.4em\relax
  IEEE, 2013, pp. 945--948.

\bibitem{liu2018rank}
Y.~Liu, X.~Yuan, J.~Suo, D.~J. Brady, and Q.~Dai, ``Rank minimization for
  snapshot compressive imaging,'' \emph{IEEE transactions on pattern analysis
  and machine intelligence}, vol.~41, no.~12, pp. 2990--3006, 2018.

\bibitem{zhang2018ffdnet}
K.~Zhang, W.~Zuo, and L.~Zhang, ``Ffdnet: Toward a fast and flexible solution
  for cnn-based image denoising,'' \emph{IEEE Transactions on Image
  Processing}, vol.~27, no.~9, pp. 4608--4622, 2018.

\bibitem{chandra2019phasepack}
R.~Chandra, T.~Goldstein, and C.~Studer, ``Phasepack: A phase retrieval
  library,'' in \emph{2019 13th International conference on Sampling Theory and
  Applications (SampTA)}.\hskip 1em plus 0.5em minus 0.4em\relax IEEE, 2019,
  pp. 1--5.

\bibitem{bian2015fourier}
L.~Bian, J.~Suo, G.~Zheng, K.~Guo, F.~Chen, and Q.~Dai, ``Fourier ptychographic
  reconstruction using wirtinger flow optimization,'' \emph{Optics express},
  vol.~23, no.~4, pp. 4856--4866, 2015.

\bibitem{mondelli2018fundamental}
M.~Mondelli and A.~Montanari, ``Fundamental limits of weak recovery with
  applications to phase retrieval,'' in \emph{Conference On Learning
  Theory}.\hskip 1em plus 0.5em minus 0.4em\relax PMLR, 2018, pp. 1445--1450.

\bibitem{agustsson2017ntire}
E.~Agustsson and R.~Timofte, ``Ntire 2017 challenge on single image
  super-resolution: Dataset and study,'' in \emph{Proceedings of the IEEE
  Conference on Computer Vision and Pattern Recognition Workshops}, 2017, pp.
  126--135.

\bibitem{bian2016fourier}
L.~Bian, J.~Suo, J.~Chung, X.~Ou, C.~Yang, F.~Chen, and Q.~Dai, ``Fourier
  ptychographic reconstruction using poisson maximum likelihood and truncated
  wirtinger gradient,'' \emph{Scientific reports}, vol.~6, p. 27384, 2016.

\bibitem{miao2015beyond}
J.~Miao, T.~Ishikawa, I.~K. Robinson, and M.~M. Murnane, ``Beyond
  crystallography: Diffractive imaging using coherent x-ray light sources,''
  \emph{Science}, vol. 348, no. 6234, pp. 530--535, 2015.

\bibitem{dabov2006image}
K.~Dabov, A.~Foi, V.~Katkovnik, and K.~Egiazarian, ``Image denoising with
  block-matching and 3d filtering,'' in \emph{Image Processing: Algorithms and
  Systems, Neural Networks, and Machine Learning}, vol. 6064.\hskip 1em plus
  0.5em minus 0.4em\relax International Society for Optics and Photonics, 2006,
  p. 606414.

\bibitem{elad2006image}
M.~Elad and M.~Aharon, ``Image denoising via sparse and redundant
  representations over learned dictionaries,'' \emph{IEEE Transactions on Image
  processing}, vol.~15, no.~12, pp. 3736--3745, 2006.

\bibitem{goldstein2009split}
T.~Goldstein and S.~Osher, ``The split bregman method for l1-regularized
  problems,'' \emph{SIAM journal on imaging sciences}, vol.~2, no.~2, pp.
  323--343, 2009.

\bibitem{zhang2017beyond}
K.~Zhang, W.~Zuo, Y.~Chen, D.~Meng, and L.~Zhang, ``Beyond a gaussian denoiser:
  Residual learning of deep cnn for image denoising,'' \emph{IEEE Transactions
  on Image Processing}, vol.~26, no.~7, pp. 3142--3155, 2017.

\bibitem{wei2020tuning}
K.~Wei, A.~Aviles-Rivero, J.~Liang, Y.~Fu, C.-B. Schnlieb, and H.~Huang,
  ``Tuning-free plug-and-play proximal algorithm for inverse imaging
  problems,'' \emph{arXiv preprint arXiv:2002.09611}, 2020.

\end{thebibliography}

\begin{addendum}
 \item This work was supported by the National Natural Science Foundation of China (Nos. 61971045, 61827901, 61991451), National Key R\&D Program (Grant No. 2020YFB0505601), Fundamental Research Funds for the Central Universities (Grant No. 3052019024).
 \item[Author contributions] Liheng Bian and Xuyang Chang conceived the idea and designed the experiments. Xuyang Chang conducted the simulations and experiments. All the authors contributed to writing and revising the manuscript, and convolved in discussions during the project.
 \item[Competing Interests] The authors declare no competing financial interests.
\end{addendum}

\newpage
\section*{Figures and tables}
\newpage

\newpage

\newpage

\newpage

\newpage

\begin{table*}[h]
  \centering  
  \fontsize{11}{9}\selectfont  
  \begin{threeparttable}  
  \caption{{{\bf Quantitative comparison under the CDI modality.}  %TWF, 
CD and KAC fail with no convergence. PLIFT and PLAMP are out of computer memory. Most of the conventional algorithms produce little improvement than AP. \emph{LPR} outperforms the other algorithms, with as much as 6dB (SNR = 30) and 0.29 (SNR = 20) imrprovement on PSNR and SSIM, respectively. We use the excess time beyond AP as the other algorithms' running time, which show that prDeep consumes the most running time. In comparison, \emph{LPR} takes the same level of running time as the conventional methods.}}  
  \label{tab:1}  
    \begin{tabular}{cccccccccc}  
    \toprule [1.6pt] 
    \multirow{2}{*}{Algorithm}&  
    \multicolumn{3}{c}{SNR=20dB}&\multicolumn{3}{c}{SNR=25dB}&\multicolumn{3}{c}{SNR=30dB}\cr  
    \cmidrule(lr){2-4} \cmidrule(lr){5-7}  \cmidrule(lr){8-10}
    &PSNR&SSIM&TIME&PSNR&SSIM&TIME&PSNR&SSIM&TIME\cr  
    \midrule  [1.6pt] 
    AP&18.46&0.50&819.67&21.75&0.58&854.37&22.29&0.65&863.14\cr  
	\cmidrule(lr){1-10}	
		%\cline{1-10}
	  WF& 19.05 & 0.52 & +27.15 &20.84& 0.62 &+31.98 &21.27&0.70&+32.41\cr  
	  %TWF& \multicolumn{3}{c}{  \ding {56}-no convergence} &\multicolumn{3}{c}{ \ding {56}-no convergence}&\multicolumn{3}{c}{ \ding {56}-no convergence}\cr  
	  RWF& 18.52 & 0.50 & +25.69 &21.98 & 0.61 &+27.53 &22.41&0.71&+27.98\cr  
    %	\cline{1-10}
	\cmidrule(lr){1-10}	
  	 AF&16.55&0.42&+28.61&19.63&0.49&+29.74&19.83&0.54&+27.29\cr  
    TAF&18.57&0.53&+26.04&21.81&0.59&+25.99&22.30&0.65&+26.49\cr  
    RAF&18.52&0.53&+22.55&21.79&0.58&+21.80&22.27&0.65&+22.19\cr  
		%\cline{1-10}
		\cmidrule(lr){1-10}	
    PLIFT& \multicolumn{3}{c}{  \ding {56}-memory limitation} &\multicolumn{3}{c}{ \ding {56}-memory limitation}&\multicolumn{3}{c}{ \ding {56}-memory limitation}\cr  
	  PLAMP& \multicolumn{3}{c}{  \ding {56}-memory limitation} &\multicolumn{3}{c}{ \ding {56}-memory limitation}&\multicolumn{3}{c}{ \ding {56}-memory limitation}\cr 
	 PMAX&16.64&0.42&+38.48&19.73&0.49&+39.04&19.97&0.54&+38.11\cr  
  	  %  \cline{1-10}	
		\cmidrule(lr){1-10}	
CD& \multicolumn{3}{c}{  \ding {56}-no convergence} &\multicolumn{3}{c}{ \ding {56}-no convergence}&\multicolumn{3}{c}{ \ding {56}-no convergence}\cr  
  %\cline{1-10}
	\cmidrule(lr){1-10}	
KAC& \multicolumn{3}{c}{  \ding {56}-no convergence} &\multicolumn{3}{c}{ \ding {56}-no convergence}&\multicolumn{3}{c}{ \ding {56}-no convergence}\cr  
	\cmidrule(lr){1-10}	
	 prDeep&20.60&0.52&+49.01&21.83&0.58&+43.36&23.33&0.65&+35.46\cr 
%\cline{1-10} 
	\cmidrule(lr){1-10}	
    LPR&{\bf 23.30}&{\bf 0.79}& +28.52&{\bf 25.52}&{\bf 0.83}&+29.97&{\bf 28.11}&{\bf 0.86}& +27.19\cr  
    \bottomrule [1.6pt]  
    \end{tabular}
    \end{threeparttable}  
\end{table*}

%\section*{Tables}
%\renewcommand{\arraystretch}{1} %控制行高  
\begin{table*}[h]
  \centering  
  \fontsize{11}{9}\selectfont  
  \begin{threeparttable}  
  \caption{{{\bf Quantitative comparison under the CDP modality (5 modulation).} The Wirtinger flow based (WF, RWF) techniques fail because of insufficient measurements. PLIFT and PLAMP are out of memory. The other methods produce little improvement or consume extremely long running time compared to AP. In comparison, \emph{LPR} consumes the same level of running time as AP, and obtains the best performance with as much as 8.3dB on PSNR (SNR = 15) and 0.61 on SSIM (SNR = 10).}}  
  \label{tab:2}  
    \begin{tabular}{cccccccccc}  
    \toprule [1.6pt] 
    \multirow{2}{*}{Algorithm}&  
    \multicolumn{3}{c}{SNR=10dB}&\multicolumn{3}{c}{SNR=15dB}&\multicolumn{3}{c}{SNR=20dB}\cr  
    \cmidrule(lr){2-4} \cmidrule(lr){5-7}  \cmidrule(lr){8-10}
    &PSNR&SSIM&TIME&PSNR&SSIM&TIME&PSNR&SSIM&TIME\cr  
    \midrule  [1.6pt] 
    AP&15.60&0.21&105.76&18.61&0.33&110.73&23.22&0.55&174.98\cr  
    
	\cmidrule(lr){1-10}	
		%\cline{1-10}
	  WF& \multicolumn{3}{c}{\ding {56}-insufficient measurements} &\multicolumn{3}{c}{\ding {56}-insufficient measurements}&\multicolumn{3}{c}{\ding {56}-insufficient measurements}\cr  
	  %TWF& \multicolumn{3}{c}{\ding {56}-insufficient measurements} &\multicolumn{3}{c}{\ding {56}-insufficient measurements}&\multicolumn{3}{c}{\ding {56}-insufficient measurements}\cr  
	  RWF& \multicolumn{3}{c}{\ding {56}-insufficient measurements} &\multicolumn{3}{c}{\ding {56}-insufficient measurements}&\multicolumn{3}{c}{\ding {56}-insufficient measurements}\cr  
    %	\cline{1-10}
	\cmidrule(lr){1-10}	
  	 AF&13.93&0.19&247.07&17.84&0.33&231.38&23.13&0.60&211.39\cr  
    TAF&13.40&0.16&257.57&18.14&0.34&225.67&22.71&0.59&213.65\cr  
    RAF&13.88&0.19&261.59&17.86&0.38&222.38&23.10&0.59&212.09\cr  
		%\cline{1-10}
		\cmidrule(lr){1-10}	
    PLIFT& \multicolumn{3}{c}{\ding {56}-memory limitation} &\multicolumn{3}{c}{\ding {56}-memory limitation}&\multicolumn{3}{c}{\ding {56}-memory limitation}\cr   
	 PLAMP& \multicolumn{3}{c}{\ding {56}-memory limitation} &\multicolumn{3}{c}{\ding {56}-memory limitation}&\multicolumn{3}{c}{\ding {56}-memory limitation}\cr  
	 PMAX&11.08&0.13&295.84&11.36&0.14&300.21&11.66&0.15&296.28\cr  
  	  %  \cline{1-10}	
		\cmidrule(lr){1-10}	
 CD&8.69&0.22&357.52&9.47&0.20&321.81&9.78&0.20&264.89\cr 
  %\cline{1-10}
	\cmidrule(lr){1-10}	
	 KAC&10.83&0.13&192.44&10.97&0.15&161.48&11.01&0.16&114.75\cr 
%\cline{1-10}
	\cmidrule(lr){1-10}	
	 prDeep&22.67&0.61&301.41&24.42&0.72&282.14&26.85&0.76&380.60\cr 
%\cline{1-10} 
	\cmidrule(lr){1-10}	
    LPR&{\bf 22.73}&{\bf 0.82}& 124.80&{\bf 26.92}&{\bf 0.88}&137.33&{\bf 31.89}&{\bf 0.94}& 228.42\cr  
    \bottomrule [1.6pt]  
    \end{tabular}  
    \end{threeparttable}  
\end{table*}

\begin{table*}[h]
  \centering  
  \fontsize{11}{8}\selectfont  
  \begin{threeparttable}
  \caption{{{\bf Quantitative comparison under the CDP modality (single modulation).} Most of the conventional algorithms fail with either no convergence or poor reconstruction quality because of extremely insufficient measurements. In comparison, \emph{LPR} still obtains the best reconstruction quality, with more than 17dB improvment on PSNR and nearly 0.8 on SSIM (SNR=20).}}  
  \label{tab:3}  
    \begin{tabular}{cccccccccc}  
    \toprule   [1.6pt] 
    \multirow{2}{*}{Algorithm}&  
    \multicolumn{3}{c}{SNR=10dB}&\multicolumn{3}{c}{SNR=15dB}&\multicolumn{3}{c}{SNR=20dB}\cr  
    \cmidrule(lr){2-4} \cmidrule(lr){5-7}  \cmidrule(lr){8-10}
    &PSNR&SSIM&TIME&PSNR&SSIM&TIME&PSNR&SSIM&TIME\cr  
    \midrule  [1.6pt]  
    AP&11.71&0.08&13.96&12.82&0.09&13.55&13.02&0.10&13.34\cr 
\cmidrule(lr){1-10}	 
    WF& \multicolumn{3}{c}{\ding {56}-insufficient measurements} &\multicolumn{3}{c}{\ding {56}-insufficient measurements}&\multicolumn{3}{c}{\ding {56}-insufficient measurements}\cr  
	  %TWF& \multicolumn{3}{c}{\ding {56}-insufficient measurements} &\multicolumn{3}{c}{\ding {56}-insufficient measurements}&\multicolumn{3}{c}{\ding {56}-insufficient measurements}\cr  
	  RWF& \multicolumn{3}{c}{\ding {56}-insufficient measurements} &\multicolumn{3}{c}{\ding {56}-insufficient measurements}&\multicolumn{3}{c}{\ding {56}-insufficient measurements}\cr 
	\cmidrule(lr){1-10}	
  	 AF&10.47&0.08&24.61&10.53&0.08&23.73&10.82&0.09&23.36\cr  
	 TAF&10.52&0.08&24.05&10.93&0.07&24.21&11.02&0.08&23.09\cr  
    RAF&10.38&0.06&26.17&10.43&0.07&25.83&10.78&0.08&25.82\cr  
\cmidrule(lr){1-10}	
PLIFT& \multicolumn{3}{c}{  \ding {56}-memory limitation} &\multicolumn{3}{c}{ \ding {56}-memory limitation}&\multicolumn{3}{c}{ \ding {56}-memory limitation}\cr  
	  PLAMP& \multicolumn{3}{c}{  \ding {56}-memory limitation} &\multicolumn{3}{c}{ \ding {56}-memory limitation}&\multicolumn{3}{c}{ \ding {56}-memory limitation}\cr 
	 PMAX&\multicolumn{3}{c}{ \ding {56}-insufficient measurements}&\multicolumn{3}{c}{ \ding {56}-insufficient measurements}&\multicolumn{3}{c}{ \ding {56}-insufficient measurements}\cr  
  	  %  \cline{1-10}	
		\cmidrule(lr){1-10}	
CD& \multicolumn{3}{c}{  \ding {56}-insufficient measurements} &\multicolumn{3}{c}{ \ding {56}-insufficient measurements}&\multicolumn{3}{c}{ \ding {56}-insufficient measurements}\cr  
  %\cline{1-10}
	\cmidrule(lr){1-10}	
KAC& \multicolumn{3}{c}{  \ding {56}-insufficient measurements} &\multicolumn{3}{c}{ \ding {56}-insufficient measurements}&\multicolumn{3}{c}{ \ding {56}-insufficient measurements}\cr  
	\cmidrule(lr){1-10}	
	 prDeep&18.29&0.39&153.41&19.21&0.54&142.34&23.92&0.68&104.84\cr  
\cmidrule(lr){1-10}	
    LPR&{\bf 21.11}&{\bf 0.81}&77.80&{\bf 25.64}&{\bf 0.87}& 81.51&{\bf 30.10}&{\bf 0.89}&62.89\cr  
	
    \bottomrule [1.6pt]   
    \end{tabular}  
    \end{threeparttable}  
\end{table*}

\end{document}